\documentclass[doublespacing]{elsart}

\usepackage{amssymb,graphicx}

\pdfoutput=1
\begin{document}
\begin{frontmatter}

\title{Quantum anticentrifugal force for wormhole geometry}

\author{Rossen Dandoloff}
\address{ Laboratoire de Physique Th\'{e}orique et
Mod\'{e}lisation , Universit\'{e} de Cergy-Pontoise,
 F-95302 Cergy-Pontoise, France}
 \ead{rossen.dandoloff@u-cergy.fr}

\begin{abstract}
We show the existence of an anticentrifugal force in a wormhole geometry in $R^3$. This counterintuitive force was shown to exist
in a flat $R^2$ space. The role the geometry plays in the appearance of this force is discussed.
\end{abstract}


\end{frontmatter}
\maketitle

Dimensionality of space plays a very important role in quantum mechanics. Quantum mechanics in flat Euclidean space in one, two and three dimensions leads to very different behavior of a quantum particle.

Especially quantum mechanics in a flat two dimensional Euclidean space $R^2$ gives very
unexpected results like e.g. the quantum anticentrifugal force for waves with zero angular momentum \cite{1} \cite{2} \cite{3}. This quantum anticentrifugal force is part of the so called
quantum fictitious forces that appear in two and three space dimensions \cite{4}. This phenomenon is due on one hand to the commutator of the radial momentum $p_r$ and the unit vector in radial direction $\frac{\vec r}{r}$ and on the other hand to the renormalization of the wave function
so that the wave function is normalized in flat space.

Constraining a particle to move on a
two-dimensional surface requires in general a special treatment for the Schr$\ddot{\mbox{o}}$dinger equation. It is not physical to suppose that the surface has no thickness at all (e.g. because of the Heisenberg uncertainty principle). If on the contrary the particle is allowed to move on a surface with finite thickness and then let the thickness go to zero, then there will appear an effective potential
 in the Schr$\ddot{\mbox{o}}$dinger equation for a particle on a curved surface: this potential depends on the mean {\bf M} and
the Gaussian curvature {\bf K} of the surface \cite{5}. If the surface is flat then there is no additional potential.

Quantum mechanics in three dimensions of a curved space so far has not
attracted much attention.
It has been shown that a spherical wave ($s$-wave) always blows up in three dimensions (with no external potential). In two dimensions, as we have already mentioned, on the contrary there is a very
counterintuitive collapsing of a $s$-wave to the origin. As the potential is attractive, in order to get localized states around the
origin of the coordinate system one has to add a Dirac-$\delta$ function potential there\cite{1}. Another more natural way to get this is to deform the
surface (creating e.g. a gaussian bump centered at the origin of the coordinate system) In this case the curvature effect of the bump
superposes with the anticentrifugal potential for $s$-waves and create a natural setup for localized states around the origin of the coordinate
system \cite{6}.

\begin{figure}[h]
\flushleft
   \includegraphics[scale=0.50]{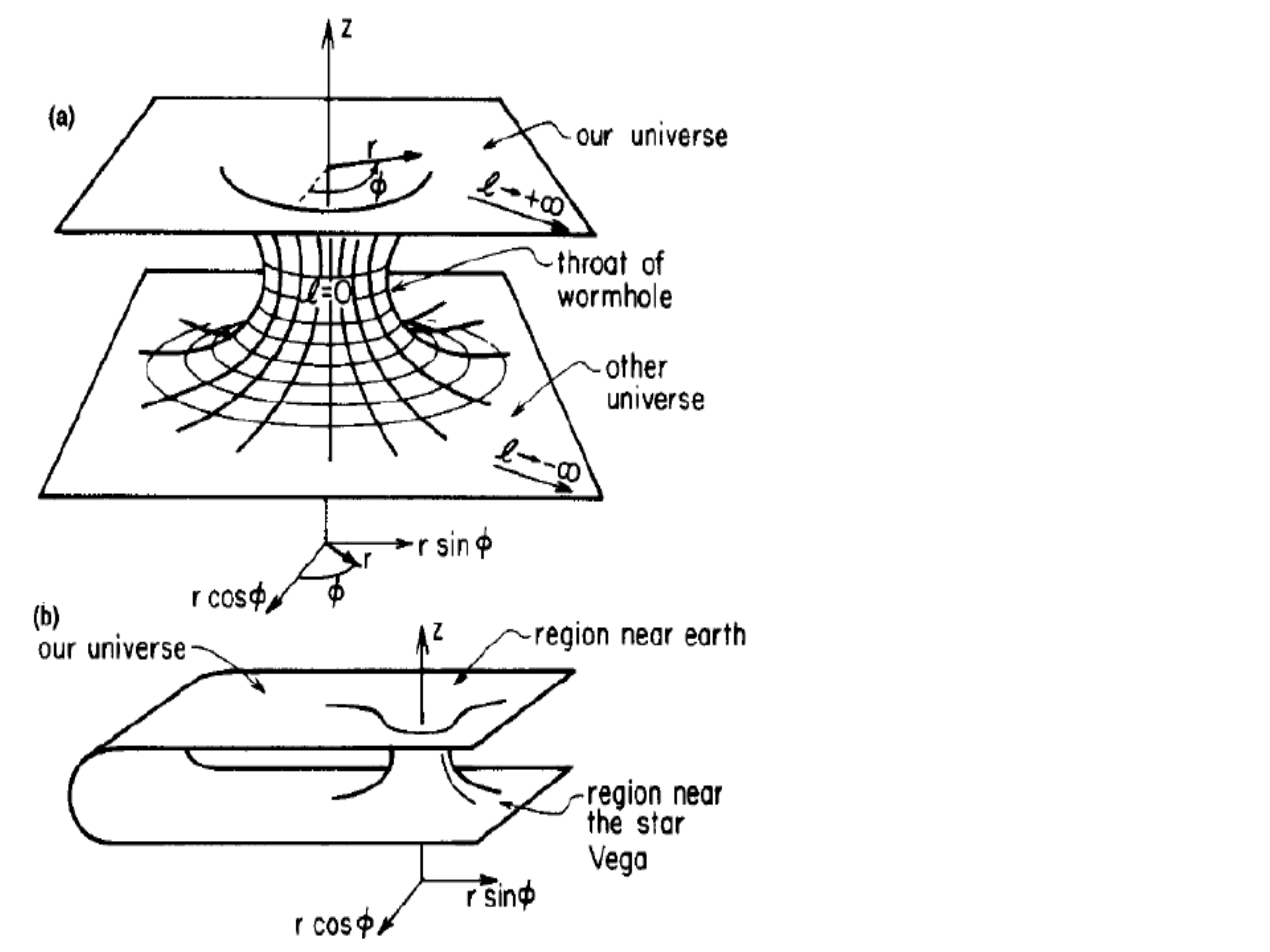}
   \caption{Schematic representation of a wormhole geometry adapted from [7]}
\end{figure}
In this paper we study the effect of curvature of a 3 dimensional space on the anticentrifugal force and show that for a suitable geometry \cite{W}
the quantum anticentrifugal potential is present in three dimensions too. This is true for $s$-waves and for higher angular momenta there are metastable states. The geometry is given by the following element of length\cite{W}:
\begin{equation}
ds^2=-c^2dt^2 + dl^2 + (b^2 + l^2)(d\theta^2+\sin^2\theta d\phi^2)
\end{equation}
  where the coordinates belong to the following intervals: $t\in[-\infty,+\infty]$, $l\in[-\infty,+\infty]$, $\theta\in[0,\pi]$ and $\phi\in[0,2\pi]$ and $b$ is the shape function of the wormhole (in general $b=b(l)$ and for $l=0, b=b(0)=b_0=const$ represents the radius of the throat of the wormhole).  $c$ is the speed of light. In this metric $t$ measures proper time of a static observer; $l$ is a radial coordinate measuring proper radial distance at constant time; $\theta$ and $\phi$ are spherical polar coordinates.
We are interested in the case $t=const$ in Eq.(1). This leads to the following metric we will use from now on:
\begin{equation}
ds^2=dl^2 + (b^2(l) + l^2)(d\theta^2+\sin^2\theta d\phi^2)\nonumber
\end{equation}
\begin{equation}
dl^2+f^2(l)d\Omega_2
\end{equation}
where $f^2(l)=(b^2(l) + l^2)$ and $d\Omega_2=d\theta^2+\sin^2\theta d\phi^2$.
Let us note here that for $b=0$ we get the usual metric for a flat $R^3$ space where $l$ plays the role of a radial coordinate. The space we are considering is shown schematically on fig.(1).

In general curvilinear coordinates the stationary Schr$\ddot{\mbox{o}}$dinger equation $-\frac{\hbar^2}{2m}\Delta\psi=E\psi$ is given by the following expression:
\begin{equation}\nonumber
 -\frac{\hbar^2}{2m}\frac{1}{h_1h_2h_3}\left[\frac{\partial}{\partial q_1}(\frac{h_2h_3}{h_1}\frac{\partial\psi}{\partial q_1})+\frac{\partial}{\partial q_2}(\frac{h_3h_1}{h_2}\frac{\partial\psi}{\partial q_2})\right.
\end{equation}

\begin{equation}
\left. + \frac{\partial}{\partial q_3}(\frac{h_1h_2}{h_3}\frac{\partial\psi}{\partial q_3}) \right]=E\psi
\end{equation}
where $h_1,\,h_2\,\,\mbox{and}\,\, h_3$ are the Lam\'e coefficients. In our case $h_1=h_l=1$, $h_2=h_\theta=\sqrt{b^2(l)+l^2}$ and $h_3=h_\phi=\sqrt{b^2(l)+l^2}\sin\theta$. The metric determinant $g=(b^2(l)+l^2)^2\sin^2\theta$. Then the Laplacian becomes:
\begin{equation}\nonumber
\Delta\psi=\frac{1}{(b^2+l^2)\sin\theta}\left[(b^2+l^2)\sin\theta\frac{\partial^2\psi}{\partial l^2}+\frac{1}{\sin\theta}\frac{\partial^2\psi}{\partial\phi^2}\right.
\end{equation}
\begin{equation}
\left.2\sin\theta(bb'+l)\frac{\partial\psi}{\partial l}+\sin\theta\frac{\partial^2\psi}{\partial\theta^2}+
\cos\theta\frac{\partial\psi}{\partial\theta}\right]
\end{equation}
In the above expression $b$ stands for $b(l)$ and $b'$ stands for $\frac{db}{dl}$. From now on we will note $b$ instead of $b(l)$. Now because we want to normalize the radial wave function with a flat norm we introduce the following ansatz as it is usually done in $R^3$:
\begin{equation}
\psi=\frac{\Phi}{\sqrt{(b^2+l^2)}}
\end{equation}
After some algebra we get the following Schr$\ddot{\mbox{o}}$dinger equation for the wave function $\Phi$:
\begin{equation}
-\frac{\hbar^2}{2m}\frac{\partial^2\Phi}{\partial l^2}+V_{eff}\Phi+\frac{1}{(b^2+l^2)}\hat{M}^2\Phi=E\Phi
\end{equation}
Where
\begin{equation}
V_{eff}=\frac{b^3b''+b^2+b'^2l^2+bb''l^2-2bb'l}{(b^2+l^2)^2}
\end{equation}
In eq.(6) $\hat{M}^2$ represents the usual angular momentum operator. Now we separate variables in "radial" and angular and set
$\Phi(l,\theta,\phi)=\Phi_1(l)\Phi_2(\theta,\phi)$. Note that as usual $\hat{M}^2\Phi_2=L(L+1)\Phi_2$. The "radial" Schr$\ddot{\mbox{o}}$dinger equation now reads:
\begin{equation}\nonumber
-\frac{\hbar^2}{2m}\frac{\partial^2\Phi_1}{\partial l^2}
+\frac{(L(L+1)}{b^2+l^2}+V_{eff}\Phi_1=E\Phi_1
\end{equation}
Let us first consider the case $b=b_0=const$. In this case the effective potential reduces to $V_{eff}=\frac{b_0^2}{(b_0^2+l^2)^2}$. This potential is repulsive and
for $b_0=0$ we recognize the usual "radial" Schr$\ddot{\mbox{o}}$dinger equation in $R^3$. For waves with angular momentum zero i.e. $L=0$ we get an overall repulsiv effective potential:
\begin{equation}
V_{eff}=\frac{\hbar^2}{2m}\frac{b_0^2}{(b_0^2+l^2)^2}
\end{equation}
contrary to what happens in two dimensional flat space $R^2$. For all other $L\not= 0$ the overall effective potential $V_{eff}$ is also repulsive.

\begin{figure}[h]
\flushleft
   \includegraphics[scale=0.30]{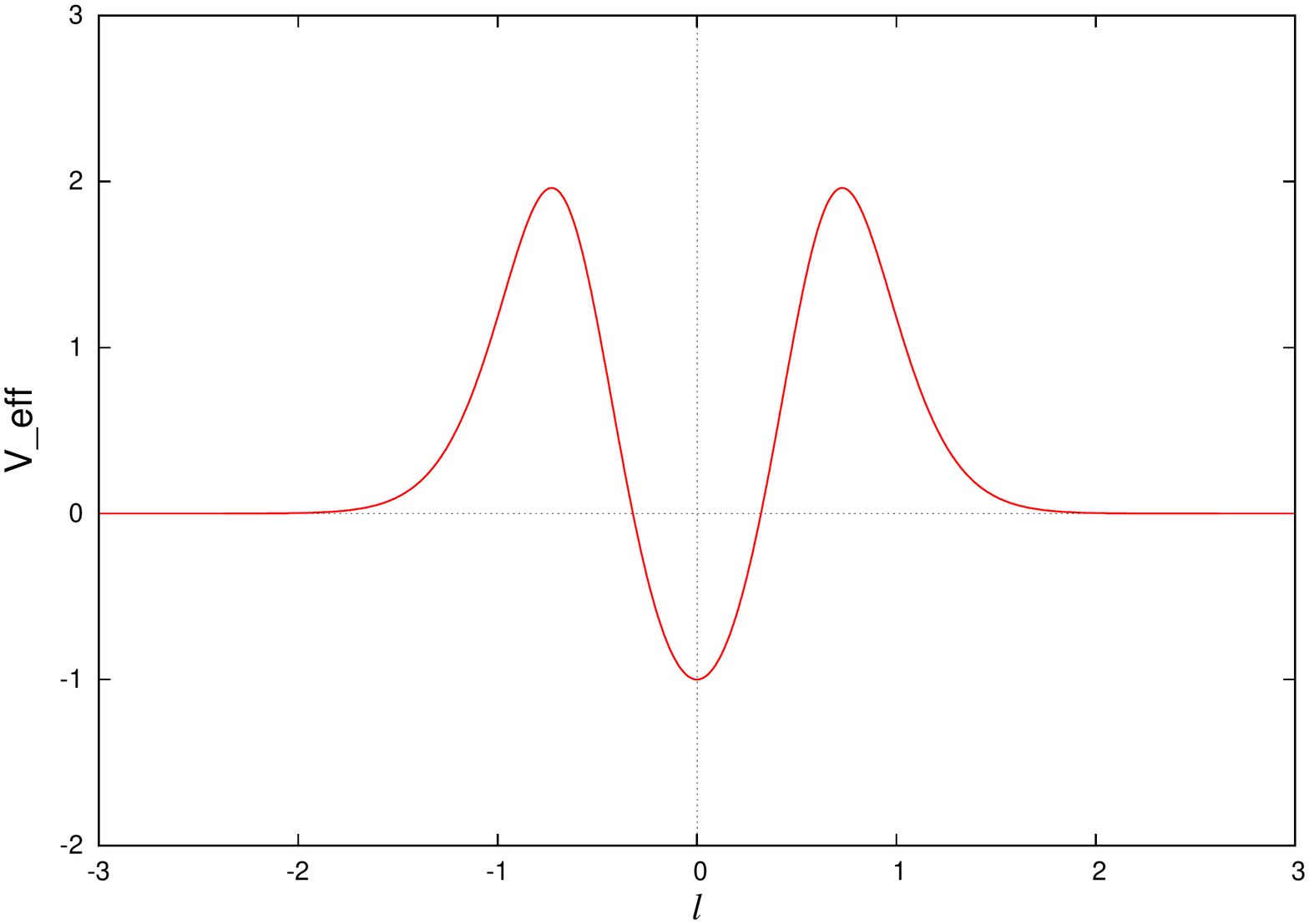}
   \caption{$V_{eff}$ for $b=b_0e^{-l^2/b_0^2}$ and L=0 (we have set $b_0=1$ in arbitrary units)}
\end{figure}
\begin{figure}[h]
\flushleft
   \includegraphics[scale=0.30]{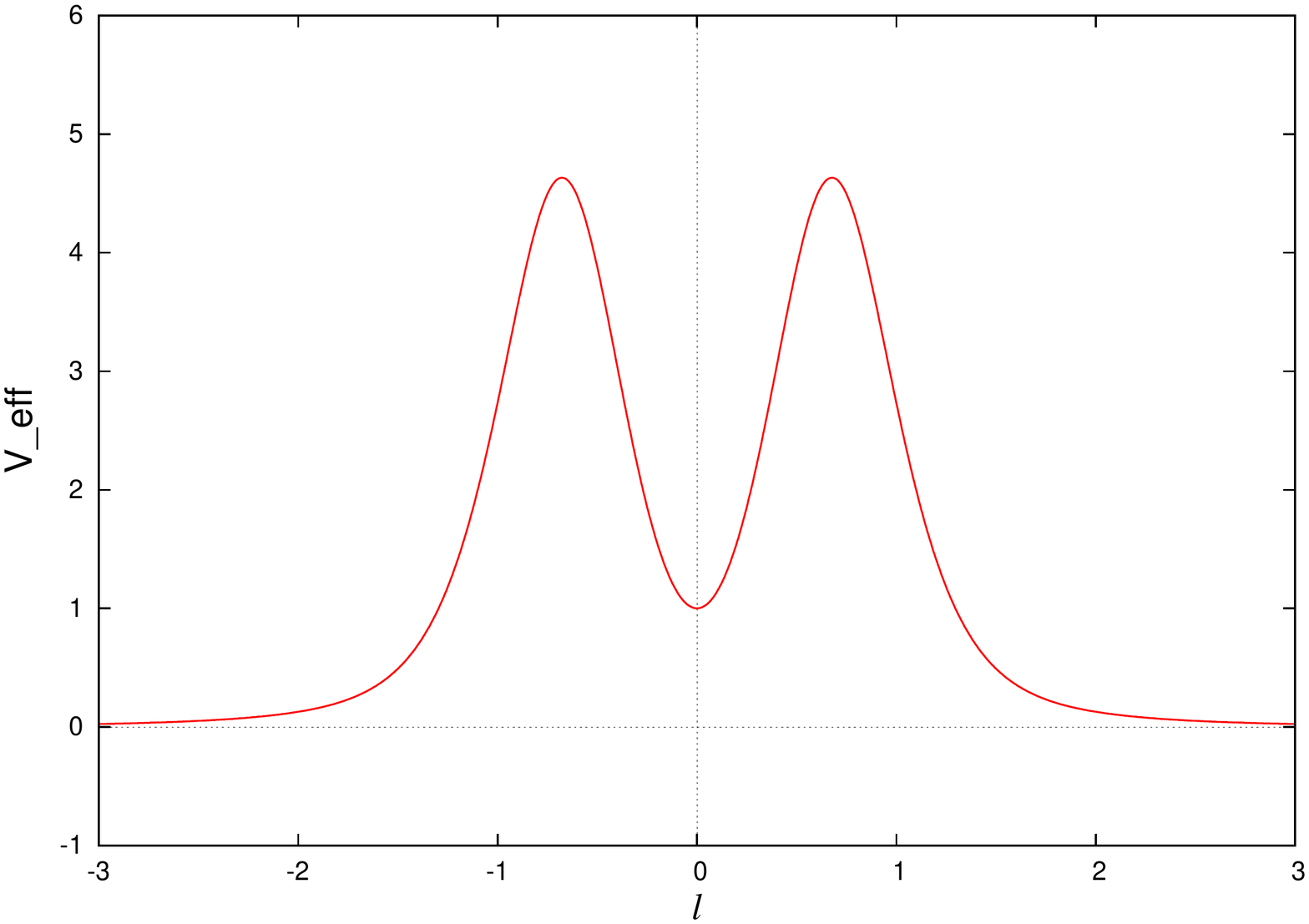}
   \caption{$V_{eff}$ for $b=b_0e^{-l^2/b_0^2}$ and L=1 (we have set $b_0=1$ in arbitrary units)}
\end{figure}

Now let us consider a more general case of a shape function $b=b_0e^{-l^2/b_0^2}$ where $b_0=const$. For very large $l$ the space is almost Euclidean
and for small $l$ the radius of the throat is almost $b_0$. Now the effective potential $V_{eff}$ has the following nontrivial form:

\begin{equation}\nonumber
V_{eff}=\frac{1}{(b_0^2e^{-2l^2/b_0^2}+l^2)^2}\left[-2b_0^2e^{-4l^2/b_0^2}+ b_0^2e^{-2l^2/b_0^2} \right.
\end{equation}

\begin{equation}
\left. +2l^2e^{-2l^2/b_0^2}+4l^2e^{-4l^2/b_0^2}
+\frac{8l^4}{b_0^2}e^{-2l^2/b_0^2}\right]
\end{equation}

This effective potential is represented schematically in Fig.2 for $b_0=1$ in arbitrary units. The depth of the potential hole at the
origin (at the center of the throat of the wormhole where $l=0$) is $V_{eff}(0)=-\frac{1}{b_0^2}$. There are obviously localized states at the
origin. Contrary to what happens in $R^2$ there is no need to add a potential at the origin in order to create bound states - here it is the geometry that creates them. In general in this case the potential is more complicated than the $R^2$ case where the potential is purely attractive.

Now, let us consider the case for $L\neq 0$. Contrary to what happens in $R^2$ or in $R^3$ for $b=b_0$ where the potential for $L\neq 0$ is purely repulsive, in this case there are
metastable states at the origin, close to the throat. In Fig.2 $V_{eff}$ is shown for angular momentum $L=1$.

It is possible to give a partial qualitative explanation of this phenomenon. As the quantum effective potential depends on the
underlying geometry it should be sensitive to any stretching of the manifold. According to the Heisenberg uncertainty relation
$\Delta q\Delta p_q =\hbar$ where $q$ is some generalized coordinate of the manifold and $p_q$ is the corresponding momentum. If in some place
the manifold is stretched $\Delta q$ is bigger than the corresponding distance between the same points in the flat embedding manifold and therefore $\Delta p_q$ is smaller and hence the energy ($E=\frac{\hbar^2(\Delta p_q)^2}{2m}$) is lower.

\begin{figure}[h]
\flushleft
   \includegraphics[scale=0.30]{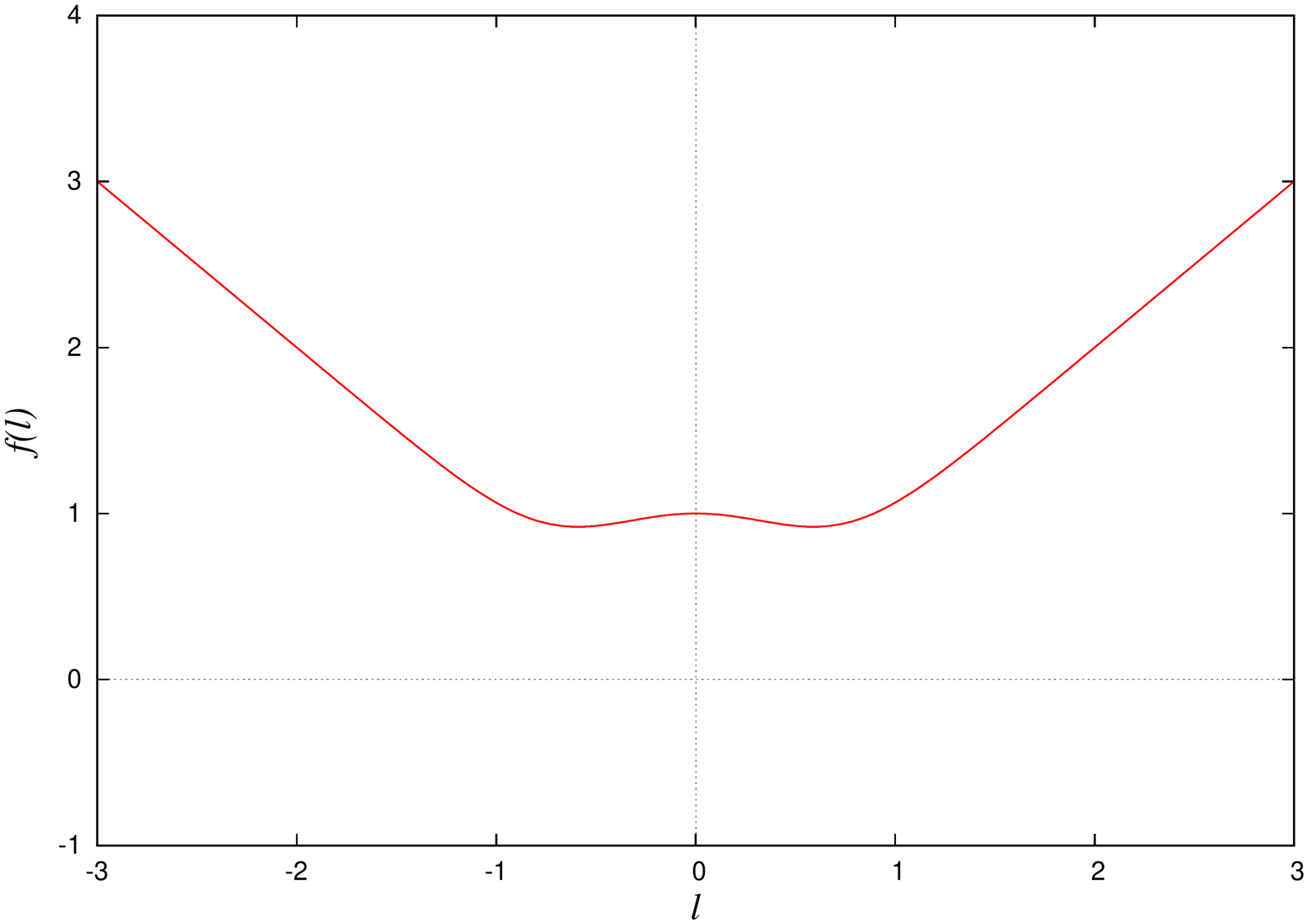}
   \caption{the "radius" f(l) of $d\Omega_2$ in eq(2) (we have set $b_0=1$ in arbitrary units)}
\end{figure}

On Fig.3 we see that
the "radius" of $d\Omega_2$ has a local maximum at the origin and obviously the manifold is "stretched" there which corresponds to a minimim
of the effective potential. Similar behavior is seen in the one and two dimensional cases \cite{6,8,9}.

We have shown that the attractive quantum effective potential can appear in three dimensions too in case of a "wormhole-geometry" when the shape function $b(l)$ has a local maximum for $l=0$. For higher
angular momenta the corresponding quantum effective potential is also nontrivial and shows metastable states at the throat of the wormhole.


\begin{thebibliography}{77}

\bibitem{1} M.A. Cirone, K. Rzazewski, W.P. Schleich, F. Straub and J.A. Wheeler, Phys. Rev. A, {\bf 65}, 022101-1, (2001)

\bibitem{2} I. Bialynicki-Birula, M.A. Cirone, J.P. Dahl, M. Fedorov and W.P. Schleich, Phys. Rev. Lett., {\bf 89}, 060404-1, (2002)

\bibitem{3} W.P. Schleich and J.P. Dahl, Phys. Rev. A, {\bf 65}, 052109, (2002);
J.P. Dahl and W.P. Schleich, Phys. Rev. A, {\bf 65}, 022109, (2002).

\bibitem{4} I. Bialynicki-Birula, M.A. Cirone, J.P. Dahl, T.H. Seligman, F. Straub and W.P. Schleich, Fortschr.Phys., {\bf 50}, 599, (2002).

\bibitem{5} R.C.T. da Costa,  Phys. Rev. A {\bf 23}, 1982 (1981).

\bibitem{6} V. Atanasov and R. Dandoloff, Phys. Lett. A, {\bf 371}, 118, (2007).

\bibitem{W} M.Morris and K.Thorn, Am. J. Phys. {\bf 56}, 395, (1988)

\bibitem{8} V.Atanasov and R. Dandoloff, Phys. Lett. A, {\bf 373}, 716, (2009); Phys.Rev. B, {\bf 79}, 033404, (2009).

\bibitem{9} R.Dandoloff and R.Balakrishnan, J.Phys. A, {\bf 38}, 6121, (2005).

\end{thebibliography}
\end{document}